\documentclass[11pt]{article}
\usepackage{Blois,epsfig,picins}

\def\Journal#1#2#3#4{{#1} {\bf #2}, #3 (#4)}

\def\NPB{{\em Nucl. Phys.} B}
\def\PLB{{\em Phys. Lett.}  B}
\def\PRL{\em Phys. Rev. Lett.}

\def\EPJC{{\em Eur. Phys. J.} C}

\def\be{\begin{equation}}
\def\ee{\end{equation}}
\def\bea{\begin{eqnarray}}
\def\eea{\end{eqnarray}}

\def\f2d3{F_2^{D(3)}}
\def\xgobs{x_\gamma^{\rm OBS}}

\def\xPom{x_{\rm I\hspace{-1.5pt}P}}
\begin{document}
\vspace*{2cm}
% \begin{center}
% \Large{\textbf{XIth International Conference on\\ Elastic and Diffractive Scattering\\ Ch\^{a}teau de Blois, France, May 15 - 20, 2005}}
% \end{center}

\vspace*{2cm}
\title{DIFFRACTIVE DIJET AND CHARM PRODUCTION AT HERA}

\author{ Y. YAMAZAKI }

\address{Institute for Particle and Nuclear Studies, KEK, Oho 1-1, 
Tsukuba, 305-0801 Japan}

\maketitle\abstracts{Jet and charm quark production in diffractive
interactions is sensitive to the partonic structure of the diffractive
exchange. This article reviews recent cross section measurements 
of such processes in both deep-inelastic scattering (DIS) regime and 
photoproduction (PHP) from the HERA $ep$ collider experiments.
The cross sections are compared to next-to-leading order QCD calculations
based on factorisation theorem.}

\section{Introduction}

A diffractive process in hadron-hadron collisions is a scattering
through a $t$-channel exchange, which has
 quantum numbers of vacuum. In $ep$ collisions,
this corresponds to a process with the diffractive exchange between 
the proton and a photon emitted from the electron. 
These processes are further classified
to diffractive photoproduction and diffractive DIS (DDIS). 
The virtuality $Q^2$ of the photon is close to zero for 
the photoproduction, while $Q^2 \gg \Lambda_{\rm QCD}$ for the DDIS. 

The cross sections of DDIS
can be described in terms of a diffractive structure function, 
$F_2^{D(4)}(\beta, Q^2, \xPom, t)$, defined as a structure function
under the condition that the reaction is a diffractive process.
Here $t$ is the four-momentum squared of the exchange at the proton vertex, 
$\xPom$ is the longitudinal momentum fraction of the diffractive exchange
to the proton and $\beta$ is defined as $\beta = x/\xPom$, where $x$ is 
the Bjorken variable in DIS. Like in inclusive DIS, a DGLAP analysis
on $F_2^{D(4)}$ allows to extract diffractive parton densities, 
$f^{D(4)}_i(z, \mu^2, \xPom, t)$, which are
defined as the parton densities of the proton, again under the condition 
that diffractive exchange occurs at $(\xPom, t)$. Here
$z$ is the longitudinal momentum fraction of the parton in the diffractive
exchange and $\mu$ is the factorisation scale.
In many mesurements $t$ cannot be reconstructed experimentally and 
the structure function integrated
over $t$, $F_2^{D(3)}$, is measured and correspondingly the parton
densities $f_i^{D(3)}$ are extracted. These parton densities are 
well constrained for quarks only since
the virtual photon in DDIS can couple directly only to quarks. The gluon
density was obtained through the scaling violation of $\f2d3$ with
much larger uncertainties (see for example \cite{h1fit2002}).

\begin{figure}
\vspace*{-3mm}
\begin{center}
\epsfig{file=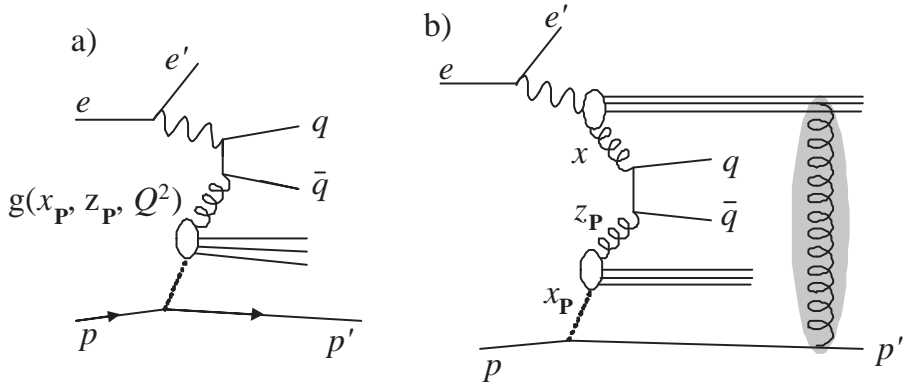,width=12cm}
\end{center}
\caption{a) An example of diagrams for dijet production
through a (virtual-)photon-gluon fusion in diffractive processes. 
b) An example of diagrams
for dijet production in resolved photon processes in diffraction
where the photon remnant has exchanged a gluon with the scattered proton.}
\label{diag}
\end{figure}

The perturbative QCD predicts that the cross sections of hard scatterings
in DDIS, such
as jet and heavy quark production, can be calculated using QCD 
factorisation\,\cite{collins}.
Namely, a
cross section at given $\xPom$ can be factorised into 
a product of hard scattering
matrix elements and the diffractive parton densities. For example,
the jet cross section $d\sigma/dE_T^{\rm jet}$ is given by
\begin{equation}
\left.\frac{d\sigma}{dE_T^{\rm jet}}\right|_{\xPom} = \sum_i\int_t dt \int_x^{\xPom}
  d\xi\frac{d\hat{\sigma}^{i\gamma^\ast}(x, \mu^2, \xi)}{dE_T^{\rm jet}}
  f_i^D(\xi, \mu^2, \xPom, t) ,
% formula for the factorisation theorem.
\end{equation}
where $\hat\sigma^{i\gamma^\ast}$ is
the parton level cross section of the photon and the $i$-th parton.
The main diagram of the jet production in
DDIS and diffractive PHP are from $\gamma^\ast g \rightarrow q\bar{q}$
($\gamma^\ast g \rightarrow Q\bar{Q}$ for heavy quark production) 
as shown in Fig.\,\ref{diag}\,a). 
Therefore cross sections for these processes are sensitive the diffractive
parton density of gluons.

\section{Factorisation tests in diffraction processes at HERA}

All the above discussions assume that QCD factorisation holds. The
measurements of $D^\ast$ meson\,\cite{zh1-disdstar} and jet\,\cite{h1disjet} 
production in DDIS
show that the cross sections are well reproduced by next-to-leading order
(NLO) calculations using the parton densities obtained from 
the HERA $\f2d3$ data. This can be interpreted in two manners; 
either the factorisation indeed holds in DDIS
assuming the parton densities used in the NLO calculations are correct, or
the data give further constraint to the parton densities assuming 
the factorisation.

\begin{figure}[t]
\centering
\begin{minipage}{7.3cm}
\vspace*{-6mm}
\begin{center}
\epsfig{file=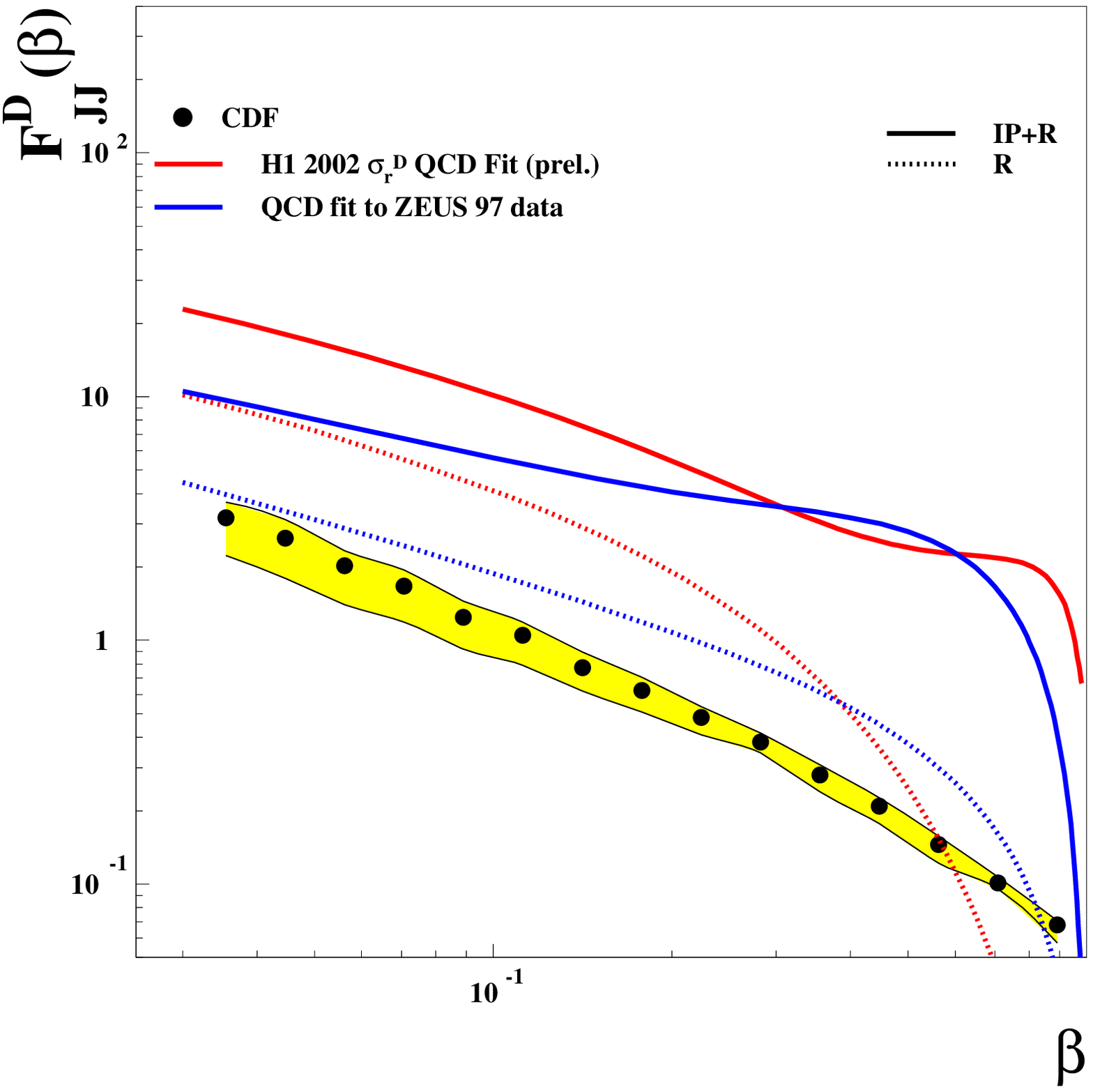,width=7.5cm}
\end{center}
\caption{
A combination of parton densities $F_{jj}^D = g(x) + (4/9)q(x)$,
extracted from  diffractive dijet cross sections measured by
CDF\,\protect\cite{cdfdij}.
The curves\,\protect\cite{heralhc} 
are the prediction using the parton densities determined
by the H1 and ZEUS data.
}
\label{tevH1Zeus}
\end{minipage}
\hspace{8mm}
\begin{minipage}{7.3cm}
\vspace*{-1mm}
\begin{center}
\epsfig{file=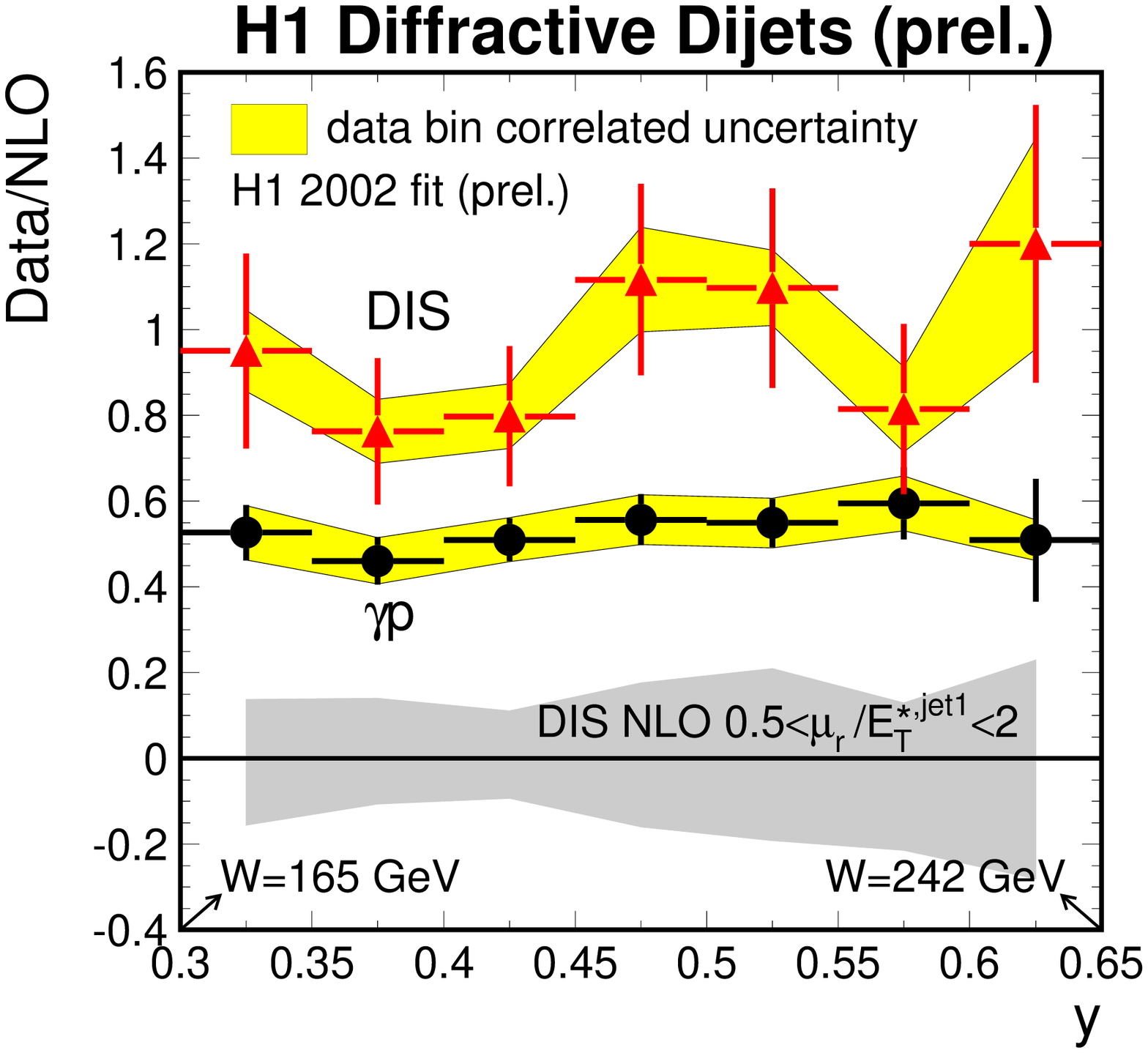,width=7.5cm}
\end{center}
\caption{The ratio of dijet cross sections to the NLO calculations,
in photoproduction and DIS diffraction\,\protect\cite{phpjeth1} as a function
of $y$. The NLO uses the H1 diffractive parton density set\,\protect\cite{h1fit2002}.}
\label{h1ratio-y}
\end{minipage}
\end{figure}

It has been known, however, that diffractive
jet production in high-energy proton-antiproton collisions at the Tevatron
is lower by factor 3--10
than perturbative QCD predictions using the HERA diffractive parton
densities as shown
in Fig.\,\ref{tevH1Zeus}. Such deficit of the cross sections is an indication
of factorisation breaking. 
A popular explanation of the suppression is multiple scattering 
in one crossing of two hadrons. 
Since the total
hadron-hadron cross section becomes larger with energy, the
probability to have more than one scatterings in one crossing becomes also
larger. Therefore, a hard scattering would often be accompanied with
peripheral scatterings, which are mostly colour-octet. This destroys
the colour-singlet state of the diffractive scattering.

Assuming that this hypothesis is correct, this should happen for all
combination of hadrons, including the hadronic component of the photon in 
photoproduction processes at HERA. That is,
the resolved photon process, where the photon resolves into more than one
partons (see Fig.\,\ref{diag}\,b)), should be suppressed. 
The direct photon process is expected to be unsuppressed as the photon
couples directly to the hard scattering (Fig.\,\ref{diag}\,a). 
Using a model explaining the suppression at the Tevatron,
Kaidalov et al. have predicted\,\cite{kaidalov} that the resolved
processes at HERA should be suppressed by a factor 0.34. 
The direct and resolved processes
can be separated using $\xgobs$, an estimator of the longitudinal
momentum fraction of the parton in the photon involved in the hard
scattering, defined as
$\xgobs = \sum(E-P_Z)_{\rm jets}/(E-P_Z)_{\rm hadrons}$.
The resolved processes are dominated in $\xgobs < 0.75$ while direct processes
are enriched in $\xgobs > 0.75$.
% The experimental signal of the resolved suppression, therefore, is that
% the events with low-$\xgobs$ in photoproduced
% diffractive events are suppressed much more than that with high-$\xgobs$ events.

The jet production in DDIS is expected to be
unsuppressed, since the photon is point-like. This is consistent with
the experimental observation described above.

\section{Recent measurements of diffractive dijet production at HERA}

The photoproduction jet cross sections have been measured recently by both the
H1\,\cite{phpjeth1} and the ZEUS\,\cite{phpjetzeus} collaborations. They are
again compared to the NLO calculations using the parton densities extracted
from the HERA data. The NLO calculations reproduced the shape of the cross
section very well. The normalisation of the observed cross sections
were, however, a factor 0.5-0.6 lower than the NLO predictions. 
Two examples of these measurements are discussed below.

The H1 collaboration has measured the jet cross sections
in DIS with a re-defined kinematical range close to the photoproduction
measurement\,\cite{phpjeth1},
in order to compare the cross sections of the DDIS and diffractive
photoproduction as directly as possible.
The measured jet cross sections for both DDIS and
diffractive photoproduction are shown as ratio to the NLO calculations using
the same diffractive parton densities (Fig.\,\ref{h1ratio-y}). 
The ratio for DDIS
is around 1, confirming that the factorisation holds using the parton densities
used in the calculation. The ratio of 
the photoproduction cross sections is about 0.5. 
This demonstrates clearly that the photoproduction cross
section is suppressed with respect to the DIS ones.

\begin{figure}
\vspace*{-8mm}
\begin{center}
\epsfig{file=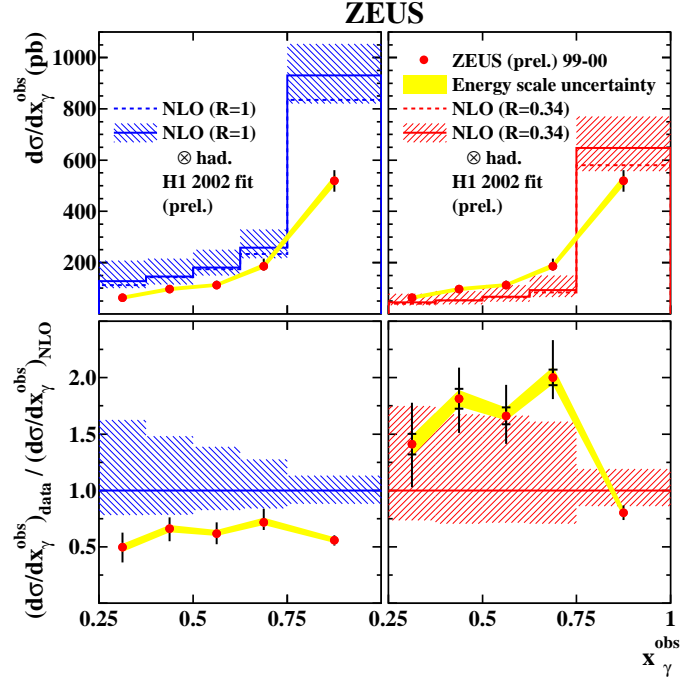,width=10.0cm}
\end{center}
\caption{The $\xgobs$ dependence of the dijet cross sections 
in photoproduction by ZEUS\,\protect\cite{phpjetzeus}. The data are compared
with the NLO calculation by Klasen et al. using the diffractive parton densities
obtained by H1\,\protect\cite{h1fit2002}, without (left) and with (right) 
the resolved suppression factor (`R' in the figure) by 
Kaidalov et al\,{\protect\cite{kaidalov}}.}
\label{zeus-xgfig}
\end{figure}

Figure\,\ref{zeus-xgfig} shows the $\xgobs$ dependence of the jet cross
sections measured by ZEUS. 
The ratio of the data to the NLO calculation is flat in $\xgobs$
at about 0.6, outside of the quoted theoretical uncertainties. The cross
section was also compared with the NLO calculations including
the resolved-only suppression by Kaidalov et al.\,\cite{kaidalov} 
The measurement excludes
this model where the cross section prediction is too sharply 
increasing with $\xgobs$. It is noted that similar observation was made in the
H1 dijet measurement mentioned above.

\section{Summary}

The measurements of jets and heavy quark production in the diffractive
processes at HERA shows that the photoproduction cross sections are lower 
by 0.5-0.6 with
respect to the NLO calculations, using the diffractive parton densities
which describe the jet production in DDIS. The suppression factor
is uniform as a function of $\xgobs$, indicating that there is no evidence
of suppression for the resolved process only, in conflict with existing
theoretical expectations. These measurements, therefore, cast a serious 
question to the current theoretical framework of the factorisation breaking
in the diffractive hadron-hadron collisions.

\end{document}